\documentclass[twocolumn,preprintnumbers,amsmath,amssymb,showpacs,prb]{revtex4}
\pdfoutput=1 % for arXiv

\usepackage{graphicx}% Include figure files
\usepackage{bm}% bold math
\usepackage{dcolumn}
\usepackage{amssymb}
\usepackage{amsmath}
\usepackage{amsfonts}
\usepackage{color}
\usepackage{epsfig}

\def \Z2{$\mathbb{Z}_2$}

\def \be{\begin{equation}}
\def \ee{\end{equation}}
\def \ba{\begin{array}}
\def \ea{\end{array}}
\def \bea{\begin{eqnarray}}
\def \eea{\end{eqnarray}}

\def \bK{{\bf K}}

\def \a{{\alpha}}

\def \D{{\Delta}}
\def \d{{\delta}}

\def \s{{\sigma}}

\begin{document}

% -------- Title -----------
\title{
    Persistence of phase boundaries between a topological and trivial $Z_2$ insulator.}
\author{
    Zohar Ringel and Ehud Altman}
\affiliation{
     Department of Condensed Matter Physics, Weizmann Institute of Science, Rehovot 76100, Israel}

% -------- Abstract -----------
\begin{abstract}
When time reversal symmetry is present there is a sharp distinction between topological and trivial band insulators which ensures that, as parameters are varied, these phases are separated by a phase transition at which the bulk gap closes. Surprisingly we find that even in the absence of time reversal symmetry, gapless regions originating from the phase boundaries persist. Moreover the critical line generically opens up to enclose Chern insulating phases that are thin but of finite extent in the phase diagram. We explain the topological origin of this effect in terms of quantized charge pumping, showing in particular that it is robust to the effect of disorder and interactions.
\end{abstract}

\maketitle

% ---- Introduction ----
\section{Introduction}

The distinction between Topological and ordinary band insulators relies on time reversal symmetry (TRS)~\cite{Kane2005,Bernevig2006}. In the presence of this symmetry the two phases must be separated in parameter space by a quantum phase transition at which the energy gap vanishes. Similarly a real space boundary between the two insulators must sustain protected metallic edge states, as has been clearly observed in experiment~\cite{Molenkamp2007,Hasan2008}. When time reversal symmetry (TRS) is broken the edge states generally become gapped or localized by disorder~\cite{Brune2010}, albeit some residual consequences of the bulk topology, such as a surface Chern number, may remain. Because there is at least a superficial relation between the edge states in real space and the gapless (critical) phase boundaries in parameter space, it is natural to ask what is the fate of the latter when time reversal symmetry is broken. Are these critical regions in the phase diagram immediately gapped out or localized, or do they contain some robust features that are not destroyed by the symmetry breaking?

In this paper we asses the robustness of the critical lines that enclose a $2D$ topological insulator in parameter space, separating it from an ordinary band insulator. We first derive general criterions for stability of those critical regions to weak TRS breaking perturbations. We further show that in certain cases the critical regions have a stronger topological protection, which does not rely on time reversal symmetry and is hence much stronger than the protection of the gapped insulators. In this case the gapless surface that enclosed the topological phase in parameter space may split up into segments to allow adiabatic connection between the former topological and ordinary insulators. However, the critical segments persist and cannot terminate for arbitrarily strong symmetry breaking perturbations.

For a two dimensional non interacting and translational invariant model, the topological criterion can be simply formulated in terms of a winding number on the Brillouin zone. The critical point between a topological and ordinary band insulator is where the bulk Dirac dispersion becomes massless. These gapless points follow a certain trajectory in $k$-space as the parameters are varied along a critical line that encloses the topological insulator in the phase diagram. We assert that the critical regions persist for arbitrary time reversal symmetry breaking, if the trajectory of Dirac points winds around the Brillouin zone.

To establish the robustness of the critical region also in presence of disorder and interactions we make a connection between the winding number on the Brillouin zone and a Chern number associated with charge pumping. Specifically the winding of the Dirac points on the Brillouin zone is directly related to a Chern number defined on a path in parameter space that encircles the topological phase from the outside, i.e. on a path that lies entirely in the ordinary insulating phase. A non zero Chern number implies pumping of an integer number of particles per row of the lattice upon following such a closed adiabatic path. Hence, the critical phase boundary might exhibit a topological protection that is much stronger than that of the insulating phases ~\cite{EhudErez}. In contrast to the latter, this topological protection does not rely on time reversal symmetry but only on particle number conservation.

%
%The structure of this work is as followed. We consider 2D, time-reversal invariant, 4-band models which can be driven to trivial and topological insulating phases by varying some parameter $\alpha,\beta...$ which respect TRS. The different topological character of the two phases implies the appearance of critical points when such a single parameter is varied, or critical lines if two parameter are varied. We start by discussing the stability of these regions, first to moderate TRS breaking perturbations and later to strong TRS-breaking perturbation and/or interactions. We then turn to a concrete example where the all the above stabilities and instabilities are realized. Last we use the tools we develop to discuss the stability of the above mention metallic phase of the ionic Hubbard model.

\section{Robust critical regions in an extended Kane and Mele model}
\label{Section-numerics}
To elucidate the main results of the paper we first demonstrate the aforementioned features on a concrete $2D$ tight binding model which can be tuned between a topological insulator (TI) and a band insulator (BI). This model is an extension of the Kane
and Mele model~\cite{Kane2005}, describing electrons hopping on a Honeycomb lattice in the presence of spin-orbit coupling. It includes two parameters, staggered hopping and staggered potential, which drive the system into a BI phase. As a function of these, the phase diagram consists of an elliptical TI region encircled by a BI phase with a critical line in between. We shall study the fate of this line as TRS is broken by a staggered and a uniform in-plane Zeeman field.

\begin{figure}[tbh]
\begin{center}
\includegraphics[trim = 50mm 20mm 40mm 10mm, clip, width=90mm]{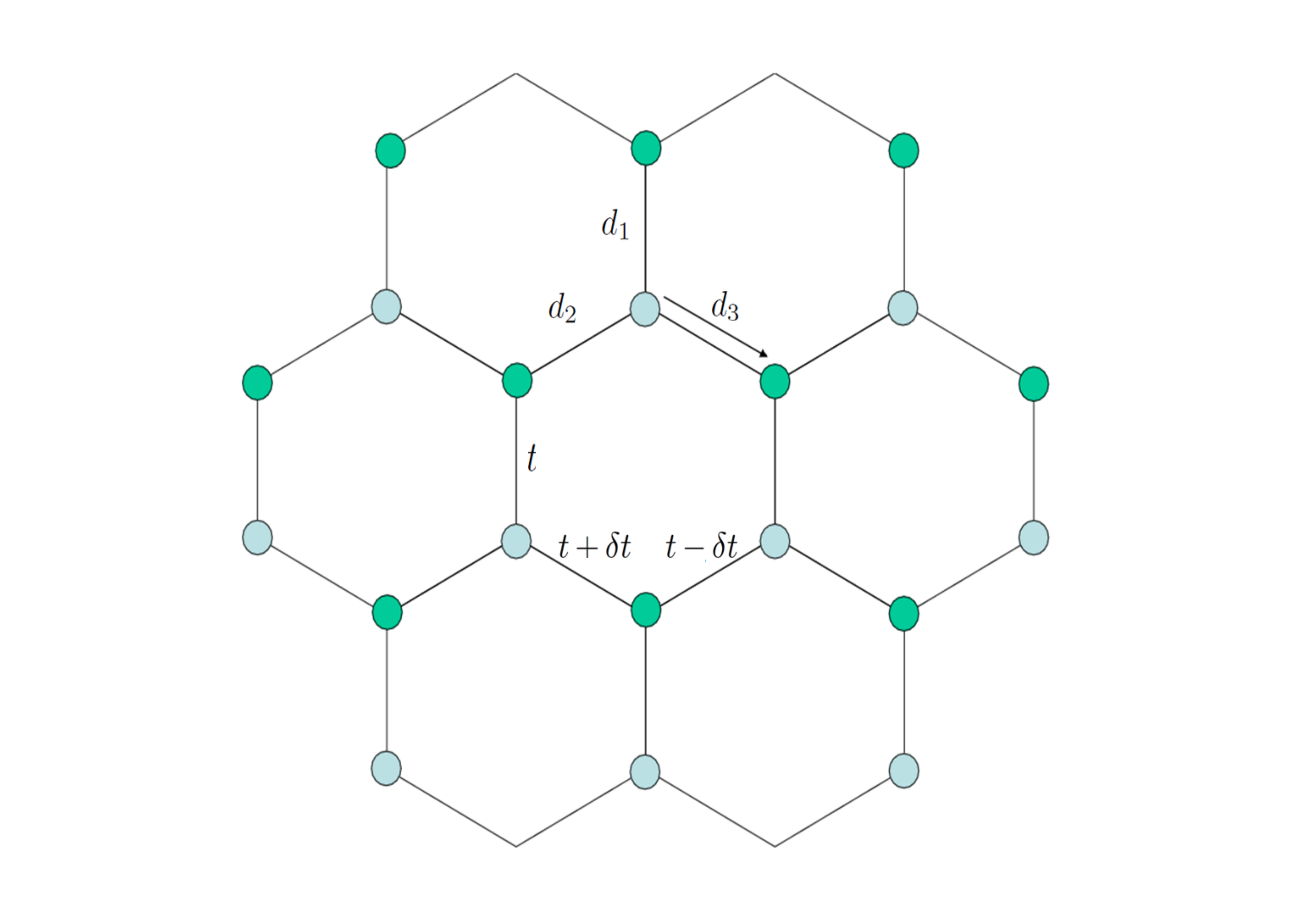}
\end{center}
\caption{Illustration of the lattice model and the different couplings given in Eq. (\ref{Eq:HHop})} \label{Fig:lattice}
\end{figure}

The corresponding Hamiltonian is given by
\begin{align}
\label{Eq:HHop}
{\rm H_{km}} &= {\rm H_{g}} + {\rm V_{so}} + {\rm V_{stag}} + {\rm V_{so}}+ {\rm V_{\delta t}}, \\ \nonumber
{\rm H_g} &= \sum_{\langle i j \rangle,s} t c^{\dagger}_{is} c_{js}, \\ \nonumber
{\rm V_{sta}} &= \Delta \sum_{is} \xi_i c^{\dagger}_{is}  c_{is}, \\ \nonumber
{\rm V_{so}} &= i\lambda_{so} \sum_{\langle \langle ij \rangle \rangle,s } \nu_{ij} s c^{\dagger}_{is}   c_{js} \\ \nonumber
&+ i \lambda_{R} \sum_{\langle ij \rangle} c^{\dagger}_{is}(\vec{\sigma}_{ss'}\times \hat{d}_{ij})_z c_{js'}. \\ \nonumber
{\rm V_{\delta t}} &=  \delta t \sum_{i-j=\pm d_3}  c^{\dagger}_{is} (1+s \lambda_R/t) c_{js} \\ \nonumber
&- \delta t \sum_{i-j=\pm d_2}  c^{\dagger}_{is}(1-s\lambda_R/t) c_{js},
\end{align}
Here $c_{is}$ annihilates an electron with $S_z = s$ on the $i'th$ site of an hexagonal lattice, $\xi_i = \pm 1$ is an alternating sign on the two sublattices, $\langle \langle ij \rangle\rangle $ denotes next-nearest-neighbor sites and $\nu_{i,j}=1(-1)$ if the lattice path between the two sites $i$ and $j$, winds clock-wise (counter clock-wise). The vectors $d_i$ and the staggered hopping pattern are shown in Fig. (\ref{Fig:lattice}). Summation of repeated indices is implicit. Notably the model includes an extra Rashba term proportional to $\delta t$, that couples to the out-of-plane spin component. The reason for adding the additional spin orbit term is to break spurious symmetries ~\cite{WhyNewRashba} of the model and thus explore generic features.

\begin{figure}[tbh]
\begin{center}
\includegraphics[width=110mm,angle=-90]{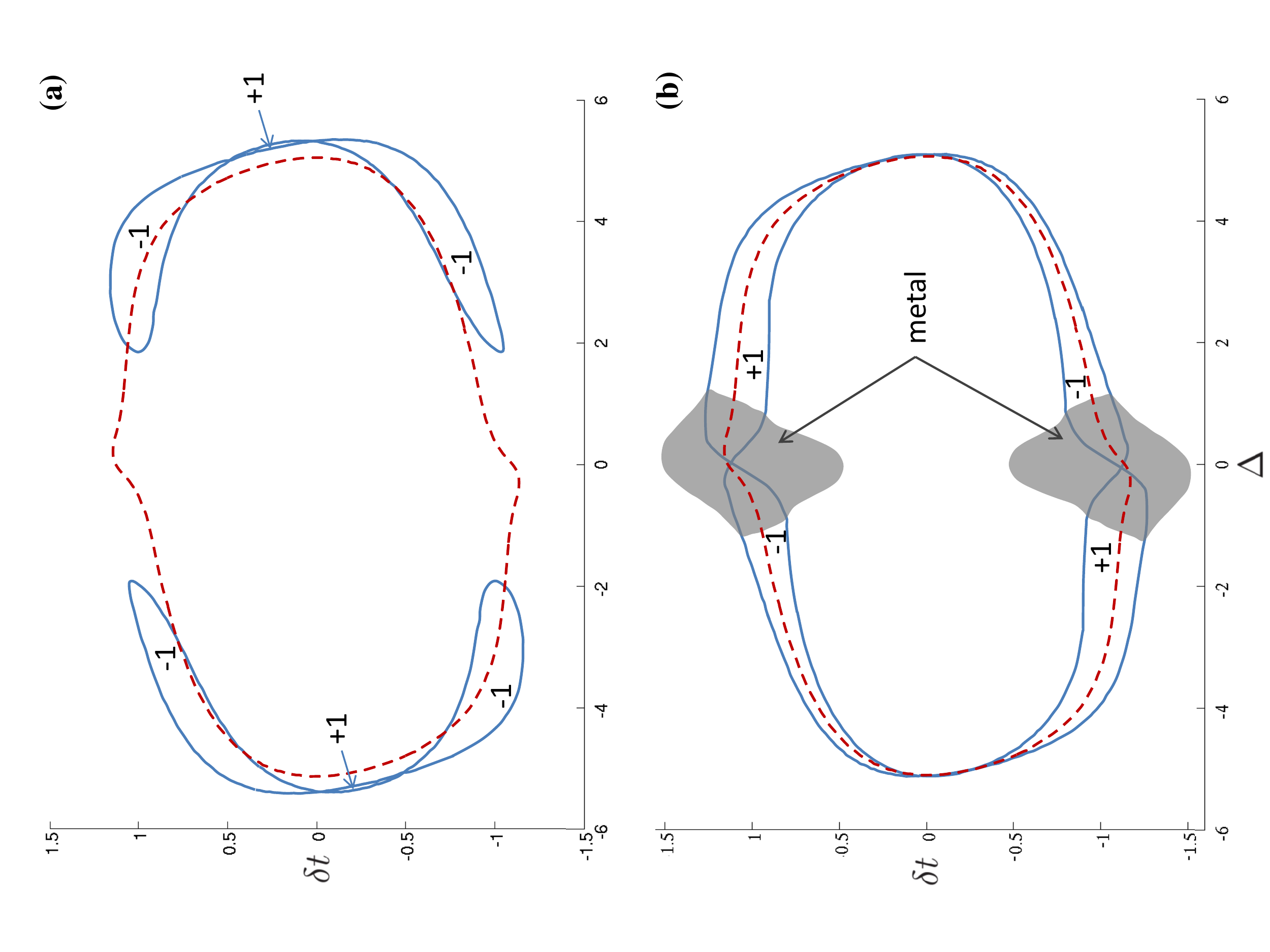}
\end{center}
\caption{Numerically obtained phase diagram of the model in Eqs. (\ref{Eq:HHop},\ref{Eq:Magnetic}) as a function of a staggered potential ($\Delta$) and staggered hopping ($\delta t$) in the presence of a staggered Zeeman field $(a)$ or a uniform Zeeman field ($b$). With TRS the ordinary and topological band insulators are separated by a line of critical points (dashed red). When time reversal symmetry is broken this line can split and generically opens up into a chern insulator phase of finite extent, having $\sigma_{xy} = \pm \frac{e^2}{h}$. Other parts of the phase boundary can open up to become metallic regions of finite extent as exemplified by the gray regions in panel (b). Notably, due to topological protection the gapless regions originating from the phase boundary persist to arbitrary large breaking of TRS.}
\label{fig:Numerics}
\end{figure}

In the absence of TRS breaking perturbations, the phase diagram of the above model in the plane of $(\Delta,\delta t)$ consists of an inner TI phase separated by a critical line from an outer BI phase, as shown in Fig. (\ref{fig:Numerics}a), dashed red line. To test the robustness of this line we consider a staggered ($M_{stg}$) and uniform ($M_{x}$) Zeeman field terms in the $x-$direction
\begin{align}
\label{Eq:Magnetic}
{\rm V_{M}} &= M_{stg} \sum_i  \xi_i c^{\dagger}_{i,\sigma} [\sigma_x]_{\sigma,\sigma'} c_{i,\sigma'} \\ \nonumber
&+ M_x \sum_i c^{\dagger}_{i,\sigma} [\sigma_x]_{\sigma,\sigma'} c_{i,\sigma'}
\end{align}
The staggered term may arise if the system breaks TRS spontaneously by developing an in-plane
anti-ferromagnetic order. We have verified that strong Hubbard-type repulsive interactions indeed lead to such a symmetry breaking with an in-plane magnetization (see also ~\cite{Karyn2010}).

Figure (\ref{fig:Numerics}) shows the numerically obtained phase diagram in $(\Delta,\delta t)$ plane for different TRS breaking perturbations and fixed values of the remaining parameters ($t=1,\lambda_{so}=1$ and $\lambda_{R} = 0.5$).  Fig. (\ref{fig:Numerics} a) and (\ref{fig:Numerics} b), show the phase diagram for non-zero staggered field ($M_{stg} = 2$) and non-zero uniform field ($M_x = 0.5$), respectively. The phase boundary between the TI and BI is marked in dashed red line. Interestingly, it does not gap out after one breaks TRS. Instead it expands into a double line (blue contour) with an IQHE phase with $\sigma_{xy}= \pm \frac{e^2}{h}$ appearing within. Metallic regions (grey regions) may also appear due to an indirect closing of the gap.

The gapless band touching lines in Fig. (\ref{fig:Numerics}) do not disappear even as we further increase the TRS breaking terms. For $M_{stg} = 5t,10t$ the eight-shaped contours in (\ref{fig:Numerics}a) simply intersect the $\Delta$ axis further away (at $\Delta \approx \pm M_{stg}$) and gradually become squeezed along this axis. For large $M_x$, the closed contour in (\ref{fig:Numerics}b) eventually splits along the $\delta t$ axis and also intersect the $\Delta$ axis near $\pm M_x$. In addition some new band touching lines appear near the original line. This apparent robustness of the band touching lines will be explained in the next two sections.

%
%Let us examine how does the oval gapless region evolve as a function of a specific TRS symmetry breaking perturbation. The perturbation we consider ($\alpha$) corresponds to $M_x = \alpha, M=0.4 \alpha$. To emphasize the features which develop, we also include an inversion symmetry breaking term which acts within the unit-cell. This term is a uniform in $k-$space and go as, $\alpha (\sigma_y \times s_x + \sqrt{2}\sigma_y \times s_y+\sqrt{3}\sigma_y \times s_z)$ where the $\sigma$ matrices acts on the sublattice indices and the $s$ matrices act on the spin indices.

%The fact that  the above results, is that the critical line typically expands into
%either a metallic phase of an Integer Quantum Hall phase. In section (\ref{Section-Perturbative}), we use an
%analogy with a Weyl Semi-metal ~\cite{Ashvin2011} to show that this is the expect for weak TRS-breaking, as long as there is no inversion symmetry. The second feature that the gapless features survive also for strong TRS-breaking is explained in section (\ref{Section-topology}) based on

\section{Weak time reversal breaking}
\label{Section-Perturbative}
The low energy theory relevant for the vicinity of the transition from an ordinary to a topological band insulator in two dimensions consists of two Dirac modes dispersing around the a pair of time reversed points K and K' in the Brillouin zone ~\cite{Onoda2007}. The effective hamiltonian for the massive Dirac fermions near one of these points is given by
\begin{align}
\label{Eq:OneConeLowE}
H_{K}(k) &= v_f \vec{k}\cdot \vec{\s}^\perp + m \sigma^z
\end{align}
where $\sigma$ acts on the space of doubly degenerate states at $K$, $\vec{k}$ measure the momenta relative to $K$ and $v_f$ is the Fermi velocity. The transition to the topological insulator is driven by changing the sign of the mass parameter $m$ to a negative value. However we have freedom to vary another parameter that changes the separation between $K$ and $K'$ while keeping $m=0$. In the extended parameter space the transition is marked by a closed critical boundary surrounding the topological phase, as shown in Fig.  (\ref{fig:Numerics}a), dashed red line. %In the case plotted in this points on the phase boundary on which the model has inversion symmetry then the two massles dirac points must meet, i.e. $K=K'$ on these points. This was the case of a critical point tuned by $\d t$ with vanishing staggered field $\D$.

\begin{figure}[tbh]
\begin{center}
\includegraphics[width=90mm]{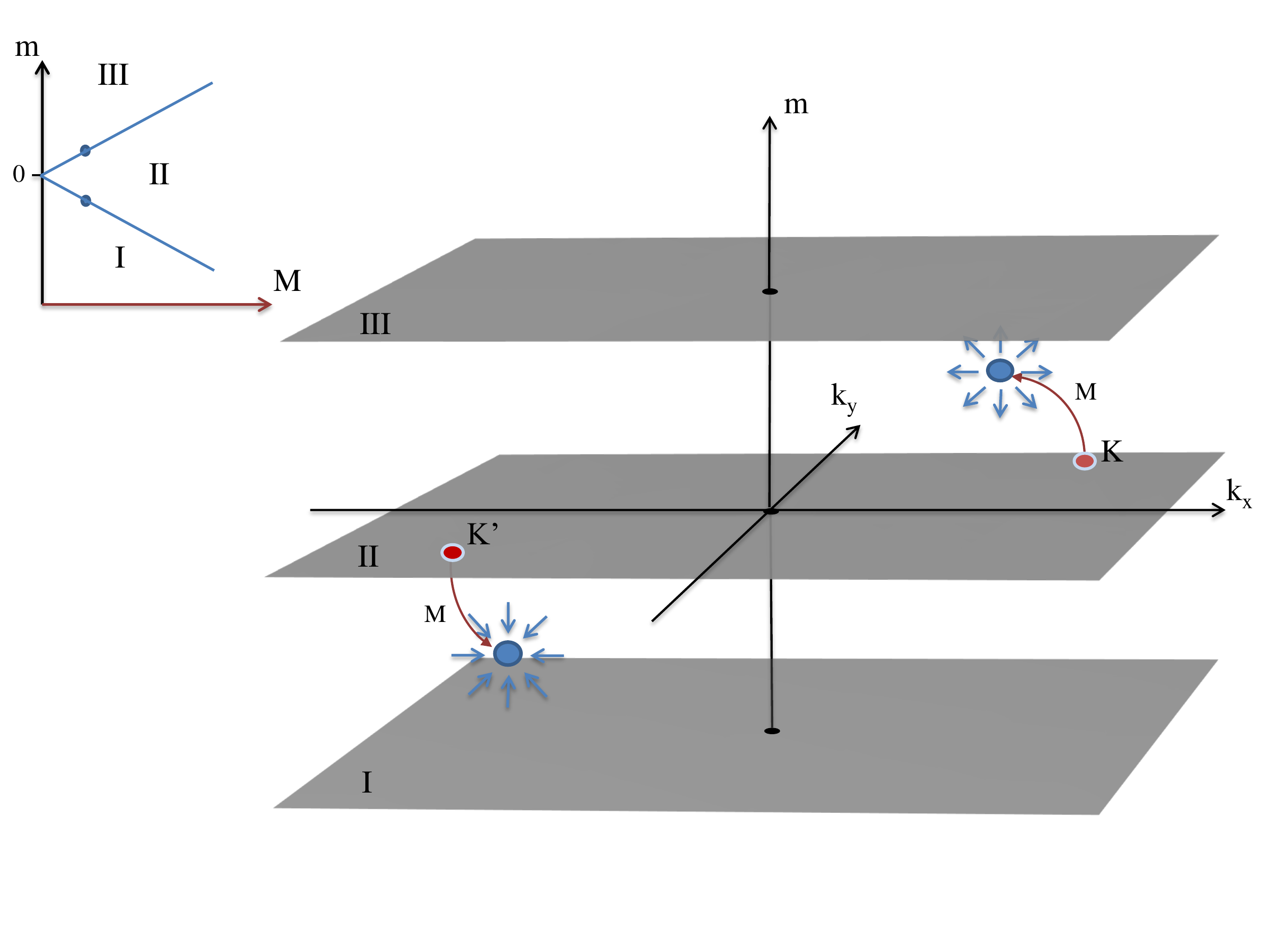}
\end{center}
\caption{Evolution of the Weyl points in the space $(k_x,k_y,m)$, where $m$ is the tuning parameter, upon breaking of TRS with a field $M$. At non zero $M$ the two Weyl points generically lie at different values of m. A system (II) with intermediate values of m is endowed with an integer Chern number from the two Weyl points.}
 \label{Fig:IQHEEmerge}
\end{figure}

It will be useful to consider the mass formally as a third spatial coordinate $k_z$. Viewed in this way, Eq. (7) describes a massless dirac cone in three dimensional space. The complete low energy theory including the two Dirac points in the extended parameter space is formally identical to a Weyl semi-metal~\cite{Ashvin2011}. In contrast to the gapless Dirac cones in two dimensional space the Dirac points in the extended three dimensional space (Weyl points) each carry a topological index that can be formulated as a Chern integer ~\cite{Berry1984}. This in particular protects the Dirac points in the extended three dimensional space from being gapped by perturbations that break time reversal symmetry as long as the $K$ and $K'$ points are spatially separated. Instead the consequence of such perturbations is to move these points away from the plane $m=0$ (see Fig. \ref{Fig:IQHEEmerge}). Moreover in absence of time reversal symmetry the Dirac points do not necessarily lie on the same $m$ coordinate

The physical space of a single system however corresponds to a plane of constant $m$. When increasing the value of $m$ with time reversal symmetry broken we generically pass two phase transitions as the plane crosses first one Dirac point and then the other.
When the plane lies either above or below both Dirac points the topological flux emanating from them cancels and the system is topologically trivial. If on the other hand the physical plane is between the the two Dirac points the topological flux adds up, endowing the system with a Chern number of $+1$ or $-1$, depending on whether the Dirac cone with a positive Chern number is above or below the plane.

The above argument implies that a small TRS breaking perturbation opens the critical phase boundary into a Chern insulating phase. The extent of this phase initially grows with increasing TRS breaking field strength. At stronger fields the Chern insulator can eventually shrink and perhaps end altogether if the Dirac cones meet and annihilate each other. Finally we should note that the two Chern bands associated with the Chern insulator may also be separated  by an indirect gap. In this case the Chern insulator would give way to a metal. This possibility is realized in certain parts of the phase diagram plotted in Fig. (\ref{fig:Numerics} b).

Before concluding this section we note that the generic situation described above can be slightly modified in presence of various spurious symmetries. Let us give a few examples relevant to the model Hamiltonian (\ref{Eq:HHop}). First in the case $\delta t=0$ the operator product of time reversal ($T$), mirror with respect to the $yz-$plane and $s_y$ is a symmetry of the Hamiltonian. As a result of this symmetry the two Dirac cones are constrained to remain on the same physical plane ($m$) even with broken TRS. For this reason the Chern insulating phases plotted in Fig. 2 are pinched at the points $\delta t=0$. Similarly if we omit the extra spin orbit term $\lambda_R$ then the product of $T$ and $s_z$ commutes with the Hamiltonian and prevents from the Chern insulator to open up at all. Furthermore, under these conditions ($\lambda_R=0$) each of the critical points on the line $\D=0$ (at $\d t=\pm 1$) must have the the Dirac points coincide $K=K'$. Then even an infinitesimal TRS breaking perturbation opens a gap for $\Delta=0$ and therefore splits the phase boundary into two separated segments.

%
%Last we comment on the possible appearance of metallic regions with a direct gap. Without a Rashba term and allowing only a staggered magnetization, the Hamiltonian anti-commutes with inversion times $\sigma_z$. This symmetry together with $TS_z$, implies an antisymmetric spectrum at each $k-$point which excludes the possibility of such metals. In the general case however, metallic regions along which the indirect gap closes may appear. This typically happens around lines where the direct gap closes as shown in the top and bottom regions of Fig. \ref{fig:2}(b).

%\begin{figure}[tbh]
%\begin{center}
%\includegraphics[width=80mm]{IQHEEmergence.png}
%\end{center}
%\caption{ } \label{Fig:IQHEEmerge}
%\end{figure}

%
%Similarly to Graphene, an effective mass model can be developed by writing the low energy electronic wave-functions as
%\begin{align}
%\Psi(r) &= [u_{Ka},u_{Kb},u_{K'a},u_{K'b}] \psi(r),
%\end{align}
%where where $u_{(K,K')(a,b)}(r)$ describe basis states at momentum $K$ and $K'$ with internal indexes $a$ or $b$ corresponding to some pseudo-spin. In terms of the vector $\psi(r)$ the Hamiltonian can be written as a 4 by 4 matrix $H(\delta k)$, where $\delta k$ is the momenta associated with $\psi(r)$.

\section{Topological stability of critical lines}
\label{Section-Topology}
In the last section we discussed the perturbative stability of the phase boundary between the topological and ordinary insulator to weak TRS fields.
However the numerical results of section (\ref{Section-numerics}) show that segments of the critical line persist to arbitrarily strong TRS breaking.
We shall argue that this is due to topological protection that is much stronger than the perturbative stability discussed above.
%We first introduce the topological invariant associated with the critical lines within the framework of non interacting band theory. We shall then show how the topological protection extends to systems with disorder and interactions by reformulation in terms of an adiabatic pumping argument.

\begin{figure}[tbh]
\begin{center}
\includegraphics[width=100mm, angle = -90]{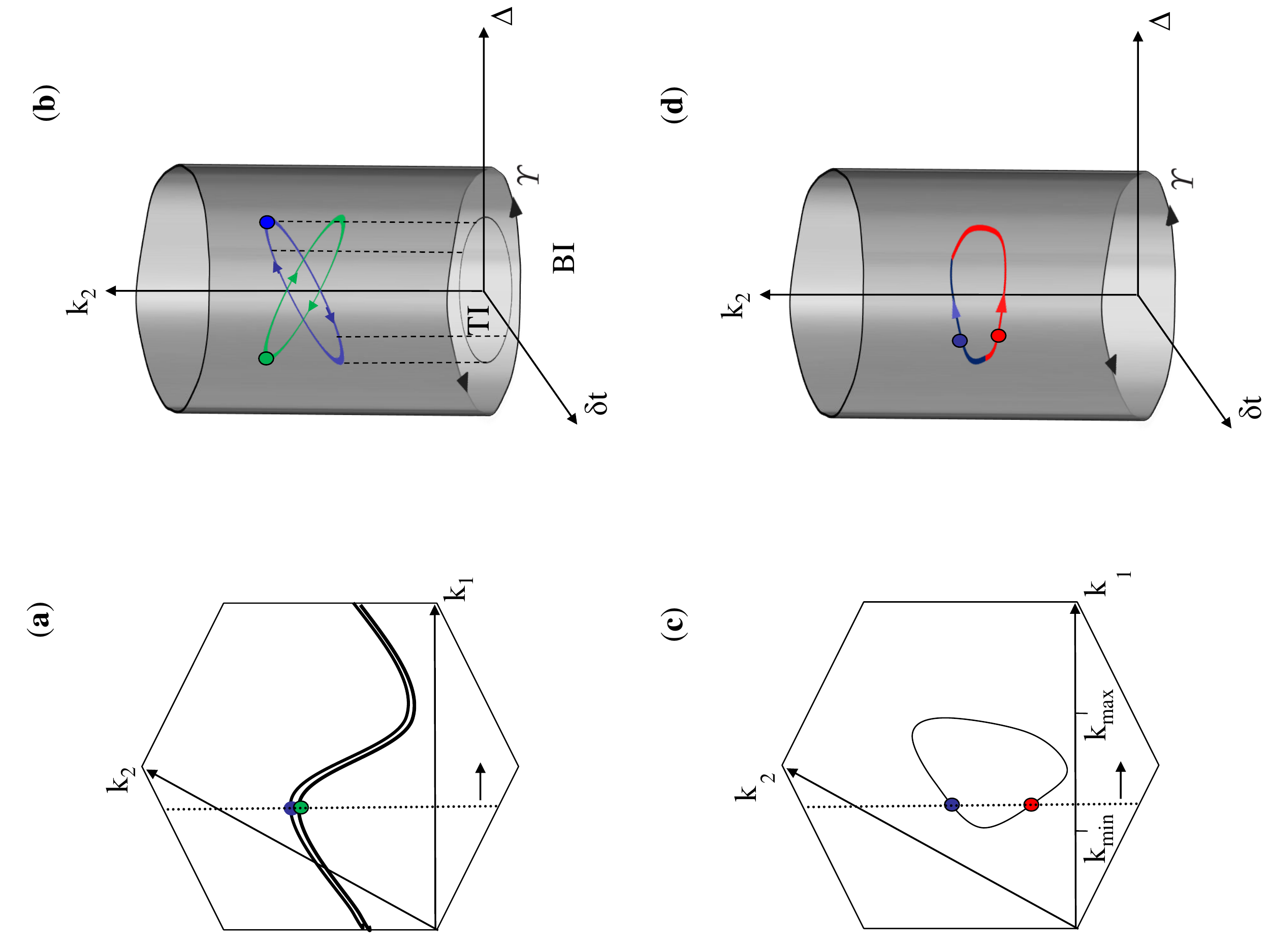}
\end{center}
\caption{The relation between windings in $k-$space and charge pumping. Panels (a) and (c) show the trajectory which the Weyl points make in the Brillouin zone if we move the system on the TI-BI phase boundary. Case (a) is an example with a topological winding number. Panels (b) and (d) show the respective trajectories of the Dirac points in the space $(\d t,\D,k_2)$ for a fixed value of $k_1$. The adiabatic path around the TI phase is extended to a distorted cylinder in that space. The  charge pumped in such a cycle is given by the topological flux on the cylinder that originates from the Weyl points. This is $\pm2$ in case (b) and zero in case (d).\label{Fig:Top}}
\end{figure}

The topological invariant is rather simple to formulate within the framework of the non interacting band theory. When TRS is broken, being on a critical line in the phase diagram is tantamount to having a single gapless Dirac cone located at some point ${\bf K}$ in the Brillouin zone. If we go along the critical line in the two parameter phase diagram, closing a full loop parameterized by $\phi\in[0,2\pi]$, the Dirac point charts at the same time a closed path ${\bf K}(\phi)$ in the two dimensional Brillouin zone. We can then associate with the critical line, as topological invariants, the two winding numbers of the loop on the toroidal Brillouin zone. Fig \ref{Fig:Top}(a) and (c) show respectively examples of a topological and a non-topological loop. In the former case the critical line must persist in the phase diagram because the Brillouin-zone loop associated with it cannot be collapsed to a point no matter how strong the time reversal breaking field.

Our goal now is to relate the winding number on the Brillouin zone to a more fundamental topological invariant associated with quantized pumping. This will allow generalization to systems with disorder and interactions where the single particle Bloch wave-functions are not eigenstates. For this purpose we consider a closed path in the two parameter space ($\D,\d t$) that encircles the topological insulator from the out-side, that is a path $\varUpsilon(\alpha)$ contained entirely in the ordinary gapped region, as depicted in Fig. (\ref{Fig:Top} b). Let us now focus on the case where the two time reversed Dirac cones (on the phase boundary) wind in the $k_1$ direction of the Brillouin zone. By fixing $k_1$ we are dealing effectively with a one dimensional Hamiltonian,
for which the pumped charge (along $k_2$) is related to an integer Chern number through the formula~\cite{Thouless,ThoulessInt}
\be
C=\sum_n \int {d\a d k_2\over \pi} {\text{Im}}\langle\partial_{\a}\psi_n|\partial_{k_2}\psi_n\rangle,
\ee
where $|\psi_n\rangle = |\psi_n(k_2,\a)\rangle$ is the wave function of the $n$-band at $k_2$ and $\a$ and the sum is over occupied bands. This is an integral over the surface of the cylinder-like surfaces shown in Fig. \ref{Fig:Top}(b,d) that measures the amount of topological flux emanating from the Dirac points enclosed by it. These Dirac points can only reside on the TI-BI phase-boundary. But for a given value of $k_1$ we are not in general guaranteed to have any. For example, if the Dirac cones at $K$ and $K'$ chart paths in $k$-space that do not wind on the Brillouin zone (see e.g. Fig \ref{Fig:Top}(c)), then we can fix $k_1$ to a value such that the effective one dimensional Hamiltonian along $k_2$ is fully gapped, even on the phase boundary. On the other hand, if the Dirac cones at $K$ and $K'$ each winds around the Brillouin zone in the direction of $k_1$ then for any fixed value of $k_1$ we must have one Dirac point on the phase boundary originating from the trajectory of $K$ and the other from $K'$.

The Dirac points originating from $K$ and from $K'$ contribute equal rather than opposite flux to the surface integral, contrary to what might have been naively guessed.
To see this first recall that each Dirac point that is enclosed by our surface can be viewed as a hedgehog singularity in the relevant three dimensional space ${\vec \kappa}\equiv (k_2,\d t,\D)$, with low energy hamiltonian near the Dirac point
$
H_{\bK,k_1}({\bf \kappa})= -\d{\vec \kappa}\cdot \vec{\s}
$.
The contribution of the Dirac point to the flux through the surface is the integer
\be
C_{\bK,k_1}=\int {d\theta d\phi \over 4\pi} \hat{\kappa}\cdot \left({\partial \hat{\kappa}\over\partial\theta }\times{\partial \hat{\kappa}\over \partial \phi}\right),
\ee
where ${\hat\kappa}={\vec\delta \kappa}/|\delta \kappa|$. This is an integer that
must remain fixed as a function of the continuous parameter $k_1$. Let us then compare the flux $C_{\bK,k_1}$ emanating from the Dirac point at $K$ to that from its time reversed partner $C_{\bK',-k_1}$. Since in the Hamiltonian $-\d{\bf \kappa}\cdot \vec{\s}$ both $k_1$ and $\s^y$, {\em and only those}, change sign under time reversal symmetry the topological flux emanating from the two time reversed points is equal. Studying the winding number of the phase boundary between the TI and BI, we found it to be $\pm1$ along the direction of the staggered hopping. Therefore adiabatically changing the parameters of the model (\ref{Eq:HHop}) to encircle the topological insulating phase, while staying in the ordinary insulator, entails pumping of exactly two electrons per row of the lattice. Such quantized pumping has been shown to be robust to disorder and interactions that invalidate the simple band picture ~\cite{ThoulessInt}. We therefore expect the gapless phase boundaries to persist even in presence of interactions .

\section{Summary and Discussion}
We showed that the critical phase boundaries separating a two dimensional topological insulator from an ordinary insulator are more robust to TRS breaking than the bulk phases themselves.  First we established the perturbative stability of the critical lines in the generic case through a mapping to a Weyl metal in one higher dimension. A corollary of this analysis is that the critical phase boundary opens up into a Chern insulating phase or into a metallic phase with finite extent in the phase diagram.
Next we formulated a criterion for yet stronger topological protection of the gapless phase boundaries that is insensitive to interactions and disorder. When the topological criterion holds the phase boundaries can split into segments, but the segments must persist to arbitrarily strong TRS breaking fields.

It is interesting to point out a surprising connection between the observations made here and previous work on the ionic Hubbard model (i.e. Hubbard model supplemented by a staggered potential). The latter is used as a model to describe transitions between a band insulator and a correlated Mott insulator with increasing interaction strength. The onset of the Mott phase is typically defined through the emergence of magnetic order, both phases are however gapped to single particle excitations. A surprising and yet unexplained observation made in a number of numerical studies is that the magnetically ordered state itself is split into two regions separated by a metallic region with vanishing single particle gap ~\cite{QMC-2,QMC-3,QMC-4}. Using a simple Hartree Fock description, the metallic phase can be seen to be smoothly connected to a phase boundary between a TI and ordinary band insulator at zero magnetization and non-vanishing spin orbit coupling (see App. \ref{App-I}). The persistence of the metallic phase deep into the magnetically ordered state, even without spin orbit, is due to the topological robustness of the phase boundary.

Finally, as an extension to this work it would be interesting to classify the robustness of phase boundaries between topological and ordinary insulators in higher dimensions (as well as higher dimensional parameter spaces). This can be done using the general classification of defects and pumps in such phases ~\cite{Rahul2011,Teo2010}.

\appendix

\section{Protected gapless regions in the ionic Hubbard model}
\label{App-I}
In this appendix we relate the metalllic regions that appear inside the magnetically ordered phase of the ionic Hubbard model ~\cite{QMC-1,QMC-2,QMC-3,QMC-4} with the persistent phase boundaries discussed in the main text of this paper. The ionic Hubbard model is defined by the following Hamiltonian on a square lattice
\begin{align}
\label{App:IHM}
{\rm H_{el}} &= \sum_{\langle ij \rangle,s} t c^{\dagger}_{i,s} c_{j,s} + \sum_{i,s} \Delta \xi_i n_{i,s} \\ \nonumber &+ \sum_{i} U (n_{i,\uparrow}-1/2) (n_{i,\downarrow}-1/2),
\end{align}
where $c_{i,s}$ annihilates an electron with spin $s$ on the $i'th$ site, $\xi_i= \pm1$ denotes an alternating sign on the two sublattices and $n_{i,s}$ is the occupancy.

The qualitative features of this model can be seen from a Hartree Fock (HF) analysis.
The interaction term is decoupled assuming a staggered magnetic moment
\begin{align}
\label{App:IHM-HF}
\sum_{i} U (n_{i,\uparrow}-1/2) (n_{i,\downarrow}-1/2) &\approx -\frac{1}{2}\sum_i M_{stg}\xi_i c^{\dagger}_{i,s} \sigma^x_{s,s'} c_{i,s'}, \\ \nonumber
\end{align}
and the resulting quadratic Hamiltonian is solved together with the self consistency condition~\cite{Footnote-Density}
\be
\vec{M}_i = \frac{U}{2} \langle c^{\dagger}_{i,s} \vec{\sigma}_{s,s'} c_{j,s'} \rangle_{\vec{M}}.
\ee

\begin{figure}
\begin{center}
\includegraphics[width=70mm, angle = 0]{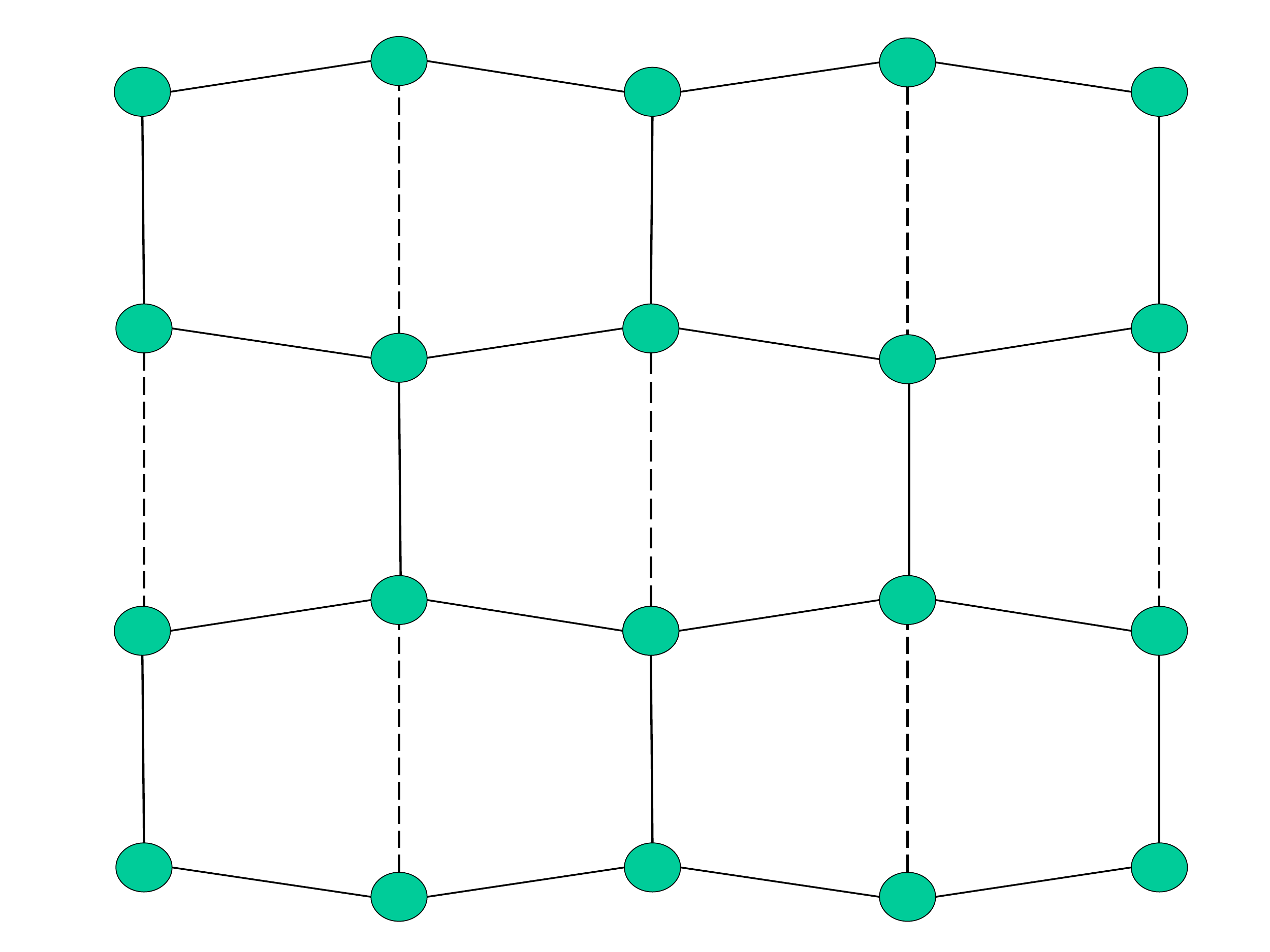}
\end{center}
\caption{A square lattice can be viewed as a deformed hexagonal lattice with one extra bond per side (dashed line).\label{Fig:Hex2Square}}
\end{figure}

The mean field theory shows a second order transition into the (AF) magnetically ordered state at a critical interaction strength $U_c \propto {t}/{\ln(t/\Delta)}$.
The single particle excitation spectrum of the mean field Hamiltonian is given by
\begin{align}
\label{App:Spec}
E_{k,s} &= \pm \sqrt{\epsilon_k^2 + (s M_{stg}  + \Delta)^2}, \\ \nonumber
\epsilon_k &= 2t(cos(k_x a) + cos(k_y a)),
\end{align}
where $s$ denotes the spin along the $x-$direction.
Notably as the magnetization increases it eventually reaches the value $|M_{stg}| = |\Delta|$. At this point the excitation gap for spins with $s = -sign(\Delta/M_{stg})$ closes and a gapless point appears.

\begin{figure}[tbh]
\begin{center}
\includegraphics[width=90mm, angle = 0]{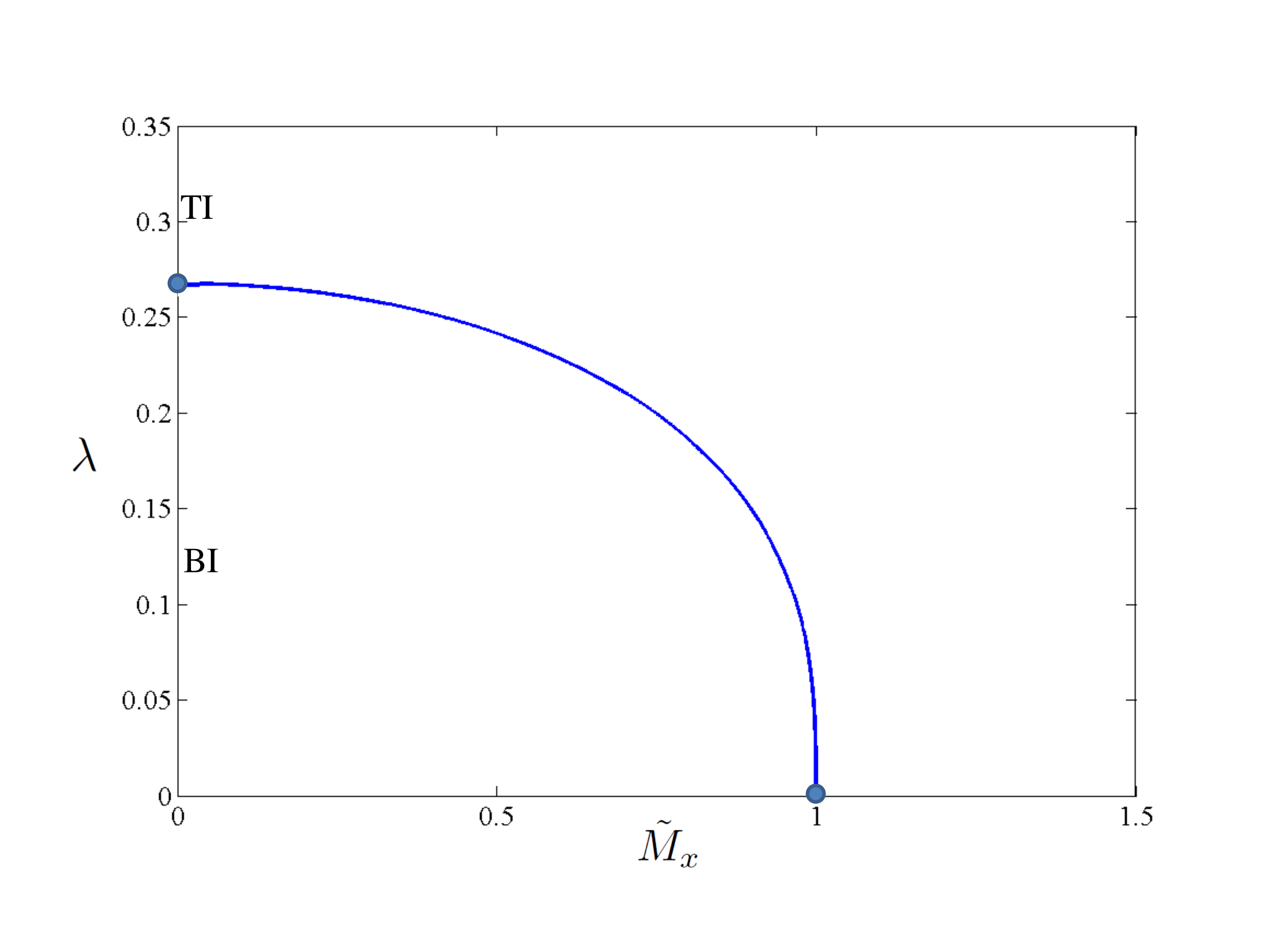}
\end{center}
\caption{Regions with gapless charge excitations in the ionic Hubbard model at $t=\Delta=1$ as a function of the spontaneous magnetization $M_{stg}$ and the spin-orbit perturbation $\lambda$ (see Eqs. (\ref{App:IHM},\ref{App:IHM-HF},\ref{App:Pert})). The $\lambda = 0$ axis corresponds to the usual ionic Hubbard model where a gapless point appears for $M_{stg}=\Delta$. Increasing $\lambda$ at $M_{stg}=0$ another gapless point appears marking the transition between a BI phase and a TI phase. Interestingly, these two points are smoothly connected implying that the former one could be seen as part of the protected gapless surface discussed in the main text. \label{Fig:ionic}}
\end{figure}

 We argue that the gapless point inside the magnetically ordered phase can be related to the critical point between a TI and BI in presence of TRS. To this end we add the following perturbation which gradually deforms the free part of ionic Hubbard model on the square lattice into the Kane and Mele model defined on a honeycomb lattice ~\cite{KaneMele}
\begin{align}
\label{App:Pert}
{\rm V} &= i\lambda \sum_{\langle \langle ij \rangle \rangle,s_z } \nu_{ij} s_z c^{\dagger}_{is_z}   c_{js_z}, \\ \nonumber
&- \lambda \sum_{\langle ij \rangle_{v},s_z} c^{\dagger}_{is_z}   c_{js_z}.
\end{align}
Here $\langle \langle ij\rangle \rangle$  denotes pairs of sites connected by two nearest neighbor bonds which do not include the dashed-bond in Fig. (\ref{Fig:Hex2Square}) and $\langle ij \rangle_{v}$ denotes sites sharing a dashed-bond. The spectrum of the mean-field Hamiltonian together with the addition ${\rm V}$ is given by
\begin{align}
\label{App:SpecSO}
E_{k,s} &= \pm \sqrt{|\epsilon_k|^2 + (s A  + \Delta)^2}, \\ \nonumber
\epsilon_k &= 2te^{i k_y a} (cos(k_x a) + cos(k_y a)) - \lambda e^{2 i k_y a}, \\ \nonumber
A &= \sqrt{4\lambda^2 [sin(2 k_x a) - 2 cos(k_y a)sin(k_x a)]^2 + M_{stg}^2},
\end{align}
where $s=\pm 1$ now denotes the spin projection along the direction $(M_{stg},0,2\lambda[sin(2 k_xa) - 2 cos(k_ya)sin(k_xa)])$. Using Eq. (\ref{App:SpecSO}) we obtain the equations for the critical points where the gap closes:
\begin{align}
\label{App:gaplessLine}
4 \lambda^2 [sin(2k_xa) - 2 sin(k_x a)]^2 &= \Delta^2 - M_{stg}^2, \\ \nonumber
cos(k_x a) = (\frac{\lambda}{2t}& - 1).
\end{align}
These equations give a critical line in the space $\lambda$ versus $M_{stg}$ which is shown in Fig. \ref{Fig:ionic}. The critical line extends all the way to the axes at $\lambda=0$ corresponding to the ionic Hubbard model on the square lattice. Thus the gapless point in the magnetically ordered phase of the ionic Hubbard model can be understood as persistence of the critical surface separating the TI and BI phases due to the topological protection discussed in the main text.

%\bibliography{AFMbib}

\begin{thebibliography}{21}
\expandafter\ifx\csname natexlab\endcsname\relax\def\natexlab#1{#1}\fi
\expandafter\ifx\csname bibnamefont\endcsname\relax
  \def\bibnamefont#1{#1}\fi
\expandafter\ifx\csname bibfnamefont\endcsname\relax
  \def\bibfnamefont#1{#1}\fi
\expandafter\ifx\csname citenamefont\endcsname\relax
  \def\citenamefont#1{#1}\fi
\expandafter\ifx\csname url\endcsname\relax
  \def\url#1{\texttt{#1}}\fi
\expandafter\ifx\csname urlprefix\endcsname\relax\def\urlprefix{URL }\fi
\providecommand{\bibinfo}[2]{#2}
\providecommand{\eprint}[2][]{\url{#2}}

\bibitem[{\citenamefont{Kane and Mele}(2005{\natexlab{a}})}]{Kane2005}
\bibinfo{author}{\bibfnamefont{C.~L.} \bibnamefont{Kane}} \bibnamefont{and}
  \bibinfo{author}{\bibfnamefont{E.~J.} \bibnamefont{Mele}},
  \bibinfo{journal}{Phys. Rev. Lett.} \textbf{\bibinfo{volume}{95}},
  \bibinfo{pages}{146802} (\bibinfo{year}{2005}{\natexlab{a}}).

\bibitem[{\citenamefont{Bernevig et~al.}(2006)\citenamefont{Bernevig, Hughes,
  and Zhang}}]{Bernevig2006}
\bibinfo{author}{\bibfnamefont{B.~A.} \bibnamefont{Bernevig}},
  \bibinfo{author}{\bibfnamefont{T.~L.} \bibnamefont{Hughes}},
  \bibnamefont{and} \bibinfo{author}{\bibfnamefont{S.-C.} \bibnamefont{Zhang}},
  \bibinfo{journal}{Science} \textbf{\bibinfo{volume}{314}},
  \bibinfo{pages}{1757} (\bibinfo{year}{2006}),
  \eprint{http://www.sciencemag.org/content/314/5806/1757.full.pdf},
  \urlprefix\url{http://www.sciencemag.org/content/314/5806/1757.abstract}.

\bibitem[{\citenamefont{K\"{o}nig et~al.}(2007)\citenamefont{K\"{o}nig,
  Wiedmann, Br\"{u}ne, Roth, Buhmann, Molenkamp, Qi, and
  Zhang}}]{Molenkamp2007}
\bibinfo{author}{\bibfnamefont{M.}~\bibnamefont{K\"{o}nig}},
  \bibinfo{author}{\bibfnamefont{S.}~\bibnamefont{Wiedmann}},
  \bibinfo{author}{\bibfnamefont{C.}~\bibnamefont{Br\"{u}ne}},
  \bibinfo{author}{\bibfnamefont{A.}~\bibnamefont{Roth}},
  \bibinfo{author}{\bibfnamefont{H.}~\bibnamefont{Buhmann}},
  \bibinfo{author}{\bibfnamefont{L.~W.} \bibnamefont{Molenkamp}},
  \bibinfo{author}{\bibfnamefont{X.-L.} \bibnamefont{Qi}}, \bibnamefont{and}
  \bibinfo{author}{\bibfnamefont{S.-C.} \bibnamefont{Zhang}},
  \bibinfo{journal}{Science} \textbf{\bibinfo{volume}{318}},
  \bibinfo{pages}{766} (\bibinfo{year}{2007}).

\bibitem[{\citenamefont{Hsieh et~al.}(2008)\citenamefont{Hsieh, Qian, Wray,
  Xia, Hor, Cava, and Hasan}}]{Hasan2008}
\bibinfo{author}{\bibfnamefont{D.}~\bibnamefont{Hsieh}},
  \bibinfo{author}{\bibfnamefont{D.}~\bibnamefont{Qian}},
  \bibinfo{author}{\bibfnamefont{L.}~\bibnamefont{Wray}},
  \bibinfo{author}{\bibfnamefont{Y.}~\bibnamefont{Xia}},
  \bibinfo{author}{\bibfnamefont{Y.~S.} \bibnamefont{Hor}},
  \bibinfo{author}{\bibfnamefont{R.~J.} \bibnamefont{Cava}}, \bibnamefont{and}
  \bibinfo{author}{\bibfnamefont{M.~Z.} \bibnamefont{Hasan}},
  \bibinfo{journal}{Nature} \textbf{\bibinfo{volume}{452}},
  \bibinfo{pages}{970} (\bibinfo{year}{2008}).

\bibitem[{\citenamefont{Br\"une et~al.}(2011)\citenamefont{Br\"une, Liu, Novik,
  Hankiewicz, Buhmann, Chen, Qi, Shen, Zhang, and Molenkamp}}]{Brune2010}
\bibinfo{author}{\bibfnamefont{C.}~\bibnamefont{Br\"une}},
  \bibinfo{author}{\bibfnamefont{C.~X.} \bibnamefont{Liu}},
  \bibinfo{author}{\bibfnamefont{E.~G.} \bibnamefont{Novik}},
  \bibinfo{author}{\bibfnamefont{E.~M.} \bibnamefont{Hankiewicz}},
  \bibinfo{author}{\bibfnamefont{H.}~\bibnamefont{Buhmann}},
  \bibinfo{author}{\bibfnamefont{Y.~L.} \bibnamefont{Chen}},
  \bibinfo{author}{\bibfnamefont{X.~L.} \bibnamefont{Qi}},
  \bibinfo{author}{\bibfnamefont{Z.~X.} \bibnamefont{Shen}},
  \bibinfo{author}{\bibfnamefont{S.~C.} \bibnamefont{Zhang}}, \bibnamefont{and}
  \bibinfo{author}{\bibfnamefont{L.~W.} \bibnamefont{Molenkamp}},
  \bibinfo{journal}{Phys. Rev. Lett.} \textbf{\bibinfo{volume}{106}},
  \bibinfo{pages}{126803} (\bibinfo{year}{2011}),
  \urlprefix\url{http://link.aps.org/doi/10.1103/PhysRevLett.106.126803}.

\bibitem[{\citenamefont{Berg et~al.}(2011)\citenamefont{Berg, Levin, and
  Altman}}]{EhudErez}
\bibinfo{author}{\bibfnamefont{E.}~\bibnamefont{Berg}},
  \bibinfo{author}{\bibfnamefont{M.}~\bibnamefont{Levin}}, \bibnamefont{and}
  \bibinfo{author}{\bibfnamefont{E.}~\bibnamefont{Altman}},
  \bibinfo{journal}{Phys. Rev. Lett.} \textbf{\bibinfo{volume}{106}},
  \bibinfo{pages}{110405} (\bibinfo{year}{2011}).

\bibitem[{Why()}]{WhyNewRashba}
\bibinfo{note}{This term is included to remove a coincidental symmetry along
  the $\Delta = 0$ line consisting of TRS times inversion times $S_z$. This
  symmery causes a metallic region to appear between the BI and TI and makes
  our analysis slightly more cumbersome. Also note that such a term will arise
  if the deformation of the hopping generates an in-plane electric field.}

\bibitem[{\citenamefont{Rachel and Le~Hur}(2010)}]{Karyn2010}
\bibinfo{author}{\bibfnamefont{S.}~\bibnamefont{Rachel}} \bibnamefont{and}
  \bibinfo{author}{\bibfnamefont{K.}~\bibnamefont{Le~Hur}},
  \bibinfo{journal}{Phys. Rev. B} \textbf{\bibinfo{volume}{82}},
  \bibinfo{pages}{075106} (\bibinfo{year}{2010}),
  \urlprefix\url{http://link.aps.org/doi/10.1103/PhysRevB.82.075106}.

\bibitem[{\citenamefont{Murakami et~al.}(2007)\citenamefont{Murakami, Iso,
  Avishai, Onoda, and Nagaosa}}]{Onoda2007}
\bibinfo{author}{\bibfnamefont{S.}~\bibnamefont{Murakami}},
  \bibinfo{author}{\bibfnamefont{S.}~\bibnamefont{Iso}},
  \bibinfo{author}{\bibfnamefont{Y.}~\bibnamefont{Avishai}},
  \bibinfo{author}{\bibfnamefont{M.}~\bibnamefont{Onoda}}, \bibnamefont{and}
  \bibinfo{author}{\bibfnamefont{N.}~\bibnamefont{Nagaosa}},
  \bibinfo{journal}{Phys. Rev. B} \textbf{\bibinfo{volume}{76}},
  \bibinfo{pages}{205304} (\bibinfo{year}{2007}),
  \urlprefix\url{http://link.aps.org/doi/10.1103/PhysRevB.76.205304}.

\bibitem[{\citenamefont{Wan et~al.}(2011)\citenamefont{Wan, Turner, Vishwanath,
  and Savrasov}}]{Ashvin2011}
\bibinfo{author}{\bibfnamefont{X.}~\bibnamefont{Wan}},
  \bibinfo{author}{\bibfnamefont{A.~M.} \bibnamefont{Turner}},
  \bibinfo{author}{\bibfnamefont{A.}~\bibnamefont{Vishwanath}},
  \bibnamefont{and} \bibinfo{author}{\bibfnamefont{S.~Y.}
  \bibnamefont{Savrasov}}, \bibinfo{journal}{Phys. Rev. B}
  \textbf{\bibinfo{volume}{83}}, \bibinfo{pages}{205101}
  (\bibinfo{year}{2011}),
  \urlprefix\url{http://link.aps.org/doi/10.1103/PhysRevB.83.205101}.

\bibitem[{\citenamefont{Berry}(1984)}]{Berry1984}
\bibinfo{author}{\bibfnamefont{M.~V.} \bibnamefont{Berry}},
  \bibinfo{journal}{Proceedings of the Royal Society of London. Series A,
  Mathematical and Physical Sciences} \textbf{\bibinfo{volume}{392}},
  \bibinfo{pages}{pp. 45} (\bibinfo{year}{1984}), ISSN
  \bibinfo{issn}{00804630},
  \urlprefix\url{http://www.jstor.org/stable/2397741}.

\bibitem[{\citenamefont{Thouless}(1983)}]{Thouless}
\bibinfo{author}{\bibfnamefont{D.~J.} \bibnamefont{Thouless}},
  \bibinfo{journal}{Phys. Rev. B} \textbf{\bibinfo{volume}{27}},
  \bibinfo{pages}{6083} (\bibinfo{year}{1983}).

\bibitem[{\citenamefont{Niu et~al.}(1985)\citenamefont{Niu, Thouless, and
  Wu}}]{ThoulessInt}
\bibinfo{author}{\bibfnamefont{Q.}~\bibnamefont{Niu}},
  \bibinfo{author}{\bibfnamefont{D.~J.} \bibnamefont{Thouless}},
  \bibnamefont{and} \bibinfo{author}{\bibfnamefont{Y.-S.} \bibnamefont{Wu}},
  \bibinfo{journal}{Phys. Rev. B} \textbf{\bibinfo{volume}{31}},
  \bibinfo{pages}{3372} (\bibinfo{year}{1985}).

\bibitem[{\citenamefont{Garg et~al.}(2006)\citenamefont{Garg, Krishnamurthy,
  and Randeria}}]{QMC-2}
\bibinfo{author}{\bibfnamefont{A.}~\bibnamefont{Garg}},
  \bibinfo{author}{\bibfnamefont{H.~R.} \bibnamefont{Krishnamurthy}},
  \bibnamefont{and} \bibinfo{author}{\bibfnamefont{M.}~\bibnamefont{Randeria}},
  \bibinfo{journal}{Phys. Rev. Lett.} \textbf{\bibinfo{volume}{97}},
  \bibinfo{pages}{046403} (\bibinfo{year}{2006}).

\bibitem[{\citenamefont{Bouadim et~al.}(2007)\citenamefont{Bouadim, Paris,
  H\'ebert, Batrouni, and Scalettar}}]{QMC-3}
\bibinfo{author}{\bibfnamefont{K.}~\bibnamefont{Bouadim}},
  \bibinfo{author}{\bibfnamefont{N.}~\bibnamefont{Paris}},
  \bibinfo{author}{\bibfnamefont{F.}~\bibnamefont{H\'ebert}},
  \bibinfo{author}{\bibfnamefont{G.~G.} \bibnamefont{Batrouni}},
  \bibnamefont{and} \bibinfo{author}{\bibfnamefont{R.~T.}
  \bibnamefont{Scalettar}}, \bibinfo{journal}{Phys. Rev. B}
  \textbf{\bibinfo{volume}{76}}, \bibinfo{pages}{085112}
  (\bibinfo{year}{2007}).

\bibitem[{\citenamefont{Paris et~al.}(2007)\citenamefont{Paris, Bouadim,
  Hebert, Batrouni, and Scalettar}}]{QMC-4}
\bibinfo{author}{\bibfnamefont{N.}~\bibnamefont{Paris}},
  \bibinfo{author}{\bibfnamefont{K.}~\bibnamefont{Bouadim}},
  \bibinfo{author}{\bibfnamefont{F.}~\bibnamefont{Hebert}},
  \bibinfo{author}{\bibfnamefont{G.~G.} \bibnamefont{Batrouni}},
  \bibnamefont{and} \bibinfo{author}{\bibfnamefont{R.~T.}
  \bibnamefont{Scalettar}}, \bibinfo{journal}{Phys. Rev. Lett.}
  \textbf{\bibinfo{volume}{98}}, \bibinfo{pages}{046403}
  (\bibinfo{year}{2007}).

\bibitem[{\citenamefont{{Roy}}(2011)}]{Rahul2011}
\bibinfo{author}{\bibfnamefont{R.}~\bibnamefont{{Roy}}},
  \bibinfo{journal}{ArXiv e-prints}  (\bibinfo{year}{2011}),
  \eprint{1104.1979}.

\bibitem[{\citenamefont{Teo and Kane}(2010)}]{Teo2010}
\bibinfo{author}{\bibfnamefont{J.~C.~Y.} \bibnamefont{Teo}} \bibnamefont{and}
  \bibinfo{author}{\bibfnamefont{C.~L.} \bibnamefont{Kane}},
  \bibinfo{journal}{Phys. Rev. B} \textbf{\bibinfo{volume}{82}},
  \bibinfo{pages}{115120} (\bibinfo{year}{2010}),
  \urlprefix\url{http://link.aps.org/doi/10.1103/PhysRevB.82.115120}.

\bibitem[{\citenamefont{Kancharla and Dagotto}(2007)}]{QMC-1}
\bibinfo{author}{\bibfnamefont{S.~S.} \bibnamefont{Kancharla}}
  \bibnamefont{and} \bibinfo{author}{\bibfnamefont{E.}~\bibnamefont{Dagotto}},
  \bibinfo{journal}{Phys. Rev. Lett.} \textbf{\bibinfo{volume}{98}},
  \bibinfo{pages}{016402} (\bibinfo{year}{2007}).

\bibitem[{Foo()}]{Footnote-Density}
\bibinfo{note}{For simplicity we ignored the density terms generated by the
  decoupling since these could be absorbed into $\Delta$ and the Chemical
  potential.}

\bibitem[{\citenamefont{Kane and Mele}(2005{\natexlab{b}})}]{KaneMele}
\bibinfo{author}{\bibfnamefont{C.~L.} \bibnamefont{Kane}} \bibnamefont{and}
  \bibinfo{author}{\bibfnamefont{E.~J.} \bibnamefont{Mele}},
  \bibinfo{journal}{Phys. Rev. Lett.} \textbf{\bibinfo{volume}{95}},
  \bibinfo{pages}{146802} (\bibinfo{year}{2005}{\natexlab{b}}).

\end{thebibliography}

\end{document}